\input harvmac
\sequentialequations
\lref\kha{K. Hashimoto, H. Hata and S. Moriyama, ``Brane Configuration from Monopole Solution in Non-Commutative Super Yang-Mills Theory'', JHEP 12 (1999) 021, hep-th/9910196. }
\lref\dba{D. Bak, ``Deformed Nahm Equation and a Noncommutative BPS 
Monopole'', Phys. Lett. B471 (1999) 149, hep-th/9910135. }
\lref\sgo{S. Goto and H. Hata, ``Noncommutative Monopole at the Second Order in $\theta$ '', Phys. Rev. D62 (2000) 085022, hep-th/0005101. }
\lref\dka{D. Kastor and E. S. Na, ``Electric Dipole Moment of a BPS Monopole'', 
Phys. Rev. D60 (1999) 025002, hep-th/9812191. }
\lref\mdo{M. R. Douglas and M. Li, ``D-Brane Realization of $N=2$ Super Yang-Mills Theory in Four Dimensions'',  hep-th/9604041. }
\lref\ddi{D. E. Diaconescu, ``D-branes, Monopoles and Nahm Equations'', 
Nucl. Phys. B503 (1997) 220, hep-th/9608163. }
\lref\igi{I. Giannakis, J. T. Liu and M. Porrati, ``Supersymmetric Sum Rules for Electromagnetic Multipoles'', Phys. Rev. D58 (1998) 045016, hep-th/9803073. }
\lref\aha{A. Hashimoto and K. Hashimoto, ``Monopoles and Dyons in Non-Commutative Geometry'', JHEP (1999) 005, hep-th/9909202. }

\lref\harvey{J. Harvey, ``Magnetic Monopoles, Duality and Supersymmetry'', in 
Trieste HEP Cosmology 1995, hep-th/9603086. }

\lref\ae{P.~C.~Aichelburg, and F.~Embacher,
``Exact superpartners of N=2 supergravity solitons,'' Phys. Rev. D34 (1986) 3006.}





\lref\aeseries{P.~C.~Aichelburg, and F.~Embacher, ``Supergravity Solitons I; General Framework,''
Phys. Rev. D37 (1988) 338; ``Supergravity Solitons II; the Free Case,'' Phys. Rev. D37 (1988) 911; 
``Supergravity Solitons III; the Background Problem,'' Phys.Rev. D37 (1988) 1436; ``Supergravity 
Solitons IV; Effective Soliton Interaction,'' Phys. Rev. D37 (1988) 2132.}

\lref\witten{E.~Witten and D.~Olive, ``Supersymmetry Algebras That Include Topological Charges,''
Phys. Lett. 78B (1978) 97.}

\lref\bogo{E.~B.~Bogomol'nyi, ``The Stability of Classical Solutions,'' Sov. J. Nucl. Phys. 24 (1976) 
389.}

\lref\jerome{J.~P.~Gauntlett, ``Low-Energy Dynamics of N=2 Supersymmetric Monopoles,''
Nucl. Phys. B411 (1994) 443, hep-th/9305068.}

\lref\jeromedual{J.~P.~Gauntlett, ``Duality and Supersymmetric Monopoles,'' 
Lectures given at 33rd Karpacz Winter School of Theoretical Physics: Duality - Strings and
Fields, hep-th/9705025.}

\lref\manton{N.~S.~Manton, ``A Remark on the Scattering of Monopoles,'' Phys. Lett. 110B (1982) 54.}

\lref\mantonlong{N.~S.~Manton, ``Monopole Interactions at Long Range,'' Phys. Lett. 154B (1985) 397.}

\lref\sen{A.~Sen, ``Dyon-Monopole Bound States, Self-Dual Harmonic Forms on the 
Multi-Monopole Moduli Space, and SL(2,Z) Invariance in String Theory,'' Phys. Lett. B329 (1994) 217.}

\lref\bktw{V.~Balasubramanian, D.~Kastor, J.~Traschen and K.~Z.~ Win, ``The Spin of the M2-Brane and Spin-Spin
Interactions via Probe Techniques,'' hep-th/9811037.}

\lref\jackiw{R.~Jackiw and C.~Rebbi, ``Solitons With Fermion Number 1/2,'' Phys. Rev. D13 (1976) 3398.}

\lref\atiyah{M.~F.~Atiyah and N.~J.~ Hitchin, ``Low-Energy Scattering of Nonabelian Monopoles,''
Phys. Lett. 107A (1985) 21.}

\lref\liu{I.~Giannakis and J.~T.~Liu, ``$N=2$ Supersymmetry and Dipole Moments,'' Phys. Rev. D58 (1998) 2509,
hep-th/9711173.}

\lref\osborn{H.~Osborn, ``Electric Dipole Moment for Supersymmetric Monopoles,'' Phys. Lett. 115B (1982) 226.}

\lref\porrati{S.~Ferrara and M.~Porrati, ``Supersymetric Sum Rules on Magnetic Dipole Moments of Arbitrary Spin
Particles,'' Phys. Lett. 288B (1992) 85.}


\Title{\vbox{\baselineskip12pt
}}
{\vbox{\centerline{\titlerm Superpartner Solutions of a BPS Monopole}
        \vskip2pt\centerline{\titlerm  in Noncommutative Space}
        }}
\bigskip
\centerline{
Euy~Soo~Na\foot{esna@theory.khu.ac.kr}  }
\bigskip
\centerline{\it Department of Physics }
\centerline{\it and Basic Sciences Research Institute,}
\centerline{\it  Kyunghee University, Seoul, 130-701, Korea }
\bigskip\bigskip
\centerline{\bf Abstract}
\medskip

We construct $U(2)$ BPS monopole superpartner solutions in $N=2$ non-commutative super Yang-Mills theory. Calculation to the second order in the noncommutative parameter $\theta$ shows that there is no electric quadrupole moment that is expected from the magnetic dipole structure of noncommutative $U(2)$ monopole. This might give an example of the nature of how supersymmetry works not changing between the commutative and noncommutative theories.  

\medskip
\Date{
}
\vfill\eject
The decoupling limit of the world volume theory on D3-branes in the NS-NS 2-form background is described by the noncommutative super Yang-Mills theory, in which BPS monopoles exist as a stable state because it can be higgsed just like the ordinary SYM.
The solution to the BPS equation of the $U(2)$ noncommutative monopole to the first order in $\theta$ 
has been studied in \kha\dba, and the second order solution is in \sgo. The solution has the generalized rotational invariance and  exhibits a dipole structure\mdo\ddi\aha in the magnetic field of the monopole\foot{This dipole structure can be visualized from the brane picture as D-string stretched between parallel 
D3-branes.  When a background $B$ field is turned on along the branes,the suspended D-string is tilted because the two endpoints carry opposite charges.}.

On the other hand, it is well known that ordinary BPS monopoles of $N=2$ Yang-Mills theory are invariant under half the
supersymmetry generators and hence form a 4-dimensional, short representation of the supersymmetry
algebra \witten \foot{See {\it e.g.} \harvey\ for a good review of this subject.}. 
From the work of  Jackiw and Rebbi \jackiw, we know that the angular momentum of spinning
monopoles is carried by the quantized states of fermionic zero-modes. For a single BPS monopole, the
fermionic zero-modes are generated by {\it infinitesimal} broken supersymmetry transformations. What we get by
acting with a {\it finite} transformation is then the backreaction of the fermionic
zero-modes on the other fields. 
Because of the quantized nature of the fermionic zero-mode states \jackiw, the fields of the monopole superpartner solution
are necessarily operator valued.


In \dka{} we 
have studied the
long-range fields of the different states in the ordinary $N=2$ BPS monopole supermultiplet.
Following  the work of Aichelburg and Embacher on $N=2$ BPS black holes \ae, we generate the fields of a monopole
``superpartner'' solution by acting on the bosonic monopole with an arbitrary, finite, broken supersymmetry
transformation. In which we have found 
that the operator valued electric dipole
moment is proportional to the angular momentum operator with a gyroelectric ratio $g=2$ and the
 quadrupole moment tensor is found to vanish identically for all spin states. 

This vanishing quadrupole moment tesor is in contrast with  the result of \ae 
{} on $N=2$ black hole superpartners for which these variations are nonzero, 
which is one of the motivation of this paper together with the fact that 
$U(1)$ part of the magnetic field of the noncommutative $U(2)$ monopole exhibits a dipole structure. If noncommutativity produces magnetic dipole structure to $U(2)$ monopole, then one can expect the electric quadrupole moment\igi {}  and  it would be interesting to see if there exists the electric quadrupole moment that is not found in ordinary $SU(2)$ monopole\dka. However, our calculation gives a negative answer. Up to $O(\theta^2)$ in tree level, we show that there is no quadrupole moment. Presumably, this result has something to do with the nature of how supersymmetry works not changing between the commutative and noncommutative theories.

In the following, we will construct the BPS monopole superpartner solutions of the $N=2$ noncommutative super Yang-Mils theory, by applying  the Seiberg-Witten map to all superpartner fields  to the second order in $\theta$. 
As a check, up to $O(\theta^2)$ we showed explicitly that the angular momentum operator and the electric dipole moment are independent of noncommutativity.

We restrict ourselves to the case where the non-vanishing component of the noncommutative parameter is 
$\theta_{12} = -\theta_{21} = \theta$, excluding the effect of time noncommutativity.
We shall take $U(2)$ as the gauge group because $SU(2)$ is not closed under the  $\star$-product that is defined by
\eqn\star{(f*g)(x) \equiv f(x)g(x) + {i\over 2}\theta^{\rho\sigma}
  \partial_{\rho}f(x)\partial_{\sigma}g(x) - {1\over8}\theta^{\rho\sigma}
\theta^{\alpha\beta}\partial_{\rho}\partial_{\alpha}f(x)\partial_{\sigma}\partial_{\beta}g(x)  + O(\theta^3).}
It replaces an ordinary multiplication in describing noncommutative theory. Small $\theta$ expansion is adopted to all fields. For example, the scalar Higgs field 
\eqn\expans{\hat S = \hat S^A T^A =(\hat S^a + \hat S^a_{(1)}
+ \hat S^a_{(2)})T^a 
+(\hat S^0 + \hat S^0_{(1)} + \hat S^0_{(2)})T^0,}
where the quantities with a hat denote those in the noncommutative description, the subscripts $(n)$ denote the quantitities at $O(\theta^n)$, $a=1,2,3$, and $T^A$ are the anti-hermitian generators of $U(2)$ Lie algebra.   Throughout this paper, this notation will be understood and other settings are the same as in \dka.

We now turn to the construction
of the noncommutative BPS monopole superpartner solutions\foot{See \dka {} for ordinary superpartner solutions in detail.}. 
We work in $N=2$ Yang-Mills theory with gauge group $U(2)$. The lagrangian is given by
\eqn\super{\eqalign{
{\cal L}_{N=2}&={\rm Tr}(-{1\over4}\hat F_{\mu\nu}*\hat F^{\mu\nu}
-{1\over4}(D_{\mu}\hat P)^2-{1\over2}(D_{\mu}\hat S)^2
-{e^2\over2}[\hat S,\hat P]^2_{*} \cr &
+i{\hat{\bar\psi}}\gamma^{\mu}* D_{\mu}\hat\psi
-e{\hat{\bar\psi}}*[\hat S,\hat \psi]_{*} 
-e{\hat{\bar\psi}}\gamma_5*[\hat P,\hat\psi]_{*}), \cr} }
where all fields are $U(2)$ Lie algebra valued, {\it e.g.} 
$\hat S=\hat S^A \hat T^A$, 
$\hat S$ and $\hat P$ are two scalar Higgs fields and $\hat \psi$ is a Dirac 
fermion.
The nonabelian electric and magnetic field strengths are defined by
$\hat E^{Ai}=-\hat F^{A0i}$ and 
$\hat B^{Ai}=-{1\over2}\epsilon^{ijk}\hat F_{jk}^A$.

Corresponding  global supersymmetry transformations  
\eqn\susyvariations{\eqalign{
\delta \hat A_{\mu}&=i\bar\alpha\gamma_{\mu}\hat\psi 
-i{\hat{\bar\psi}}\gamma_{\mu}\alpha,\qquad
\delta \hat P=\bar\alpha\gamma_5\hat\psi
-{\hat{\bar\psi}}\gamma_5\alpha, \qquad
\delta \hat S=i\bar\alpha\hat\psi
-i{\hat{\bar\psi}}\alpha, \cr
\delta\hat\psi&=(\half\gamma^{\mu\nu}\hat F_{\mu\nu}
-i\gamma^\mu D_\mu \hat S +\gamma^\mu D_\mu \hat P\gamma_5
-i[\hat P,\hat S]_{*} \gamma_5)\alpha, \cr }}
where the parameter $\alpha$ is a Grassmann valued Dirac spinor\foot{Our conventions for the 
Minkowski metric are ``mostly minus''
$\eta_{\mu\nu}={\rm diag}(+1,-1,-1,-1)$ and $\gamma_5=+i\gamma_0\gamma_1\gamma_2\gamma_3$.}. 
For a static, BPS monopole field configuration with 
$\hat P=\hat A_0=\hat \psi=0$ 
and 
\eqn\bps{D_i\hat S^A =\half\epsilon_{ijk}\hat F_{jk}^A,}
only the fermion $\hat \psi$ has a nontrivial supersymmetry variation given by
\eqn\fermionvar{\eqalign{\delta\hat\psi^a&=-2\gamma^k(D_k\hat S^a)P_-\alpha,\cr
\delta\hat\psi_{(1)}^0&=-2\gamma^k(D_k\hat S)_{(1)}^0P_-\alpha,\cr
\delta\hat\psi_{(2)}^a&=-2\gamma^k(D_k\hat S)_{(2)}^a P_-\alpha,\cr
}}
where 
$P_\pm=\half(1\pm\Gamma_5)$ are projection operators with $\Gamma_5=-i\gamma_0\gamma_5$.
If we define projected spinors $\alpha_\pm$ satisfying $P_\pm\alpha_\pm=\alpha_\pm$, then $\alpha_+$
generates unbroken supersymmetry transformations, while $\alpha_-$ generates broken supersymmetry transformations.
The variation $\delta\psi$ under a broken supersymmetry transformation gives a zero-mode of the fermion field
equation in the monopole background.

At second order variations, $\delta^2S$ and $\delta^2A_k$ vanish and
we find only nonzero variations for $P$ and $A_0$ given by
\eqn\secondorder{\eqalign{
\delta^2 A_0^a&=-\delta^2P^a= -4i\left(\alpha^\dagger\gamma^k\alpha\right)
D_k S^a,\cr
 (\delta^2 A)_{(1)}^0&=-(\delta^2P)_{(1)}^0= 
-4i\left(\alpha^\dagger\gamma^k\alpha\right)(D_k S)_{(1)}^0,\cr
 (\delta^2 A_0)_{(2)}^a&=-(\delta^2P)_{(2)}^a= 
-4i\left(\alpha^\dagger\gamma^k\alpha\right)(D_k S)_{(2)}^a.\cr
}}
These reduce to dipole fields in the long range limit. Interestingly, the third and
fourth order variations of all the fields turn out to vanish even in the noncommutative sector. In particular, the third order variation of
$\psi$ is found to be 
\eqn\thirdorder{\eqalign{
{\delta^3\psi^a}&=8i\left(\alpha^\dagger\gamma^k\alpha\right)
\{{\gamma^0\gamma^l}{D_lD_kS^a} 
+e{\gamma^0}{\epsilon}^{abc}(D_kS^b)S^c \}P_+\alpha,\cr
\delta^3\psi_{(1)}^0&=8i\left(\alpha^\dagger\gamma^k\alpha\right)
\{\gamma^0\gamma^l\{\partial_lD_kS_{(1)}^0
-{1\over 2}{\theta^{\rho\sigma}}\partial_\rho (D_kS^a)\partial_\sigma A_l^a\}\cr &
\qquad +e\gamma^0{1\over 2}{\theta^{\rho\sigma}}\partial_\rho (D_kS^a)
\partial_\sigma S^a \} P_+\alpha,\cr
\delta^3\psi_{(2)}^a&=8i\left(\alpha^\dagger\gamma^k\alpha\right)
\{\gamma^0\gamma^l\,
 [\,\partial_lD_kS_{(2)}^a \cr &
\qquad -e\epsilon^{abc}((D_kS^b)A_{l(2)}^c
-{1\over8}\theta^{\rho\sigma}\theta^{\alpha\beta}
\partial_\rho\partial_\alpha(D_kS^b)\partial_\sigma \partial_\beta A_i^c
 ) \cr &
\qquad\qquad -{1\over2}{\theta^{\rho\sigma}}
 (\partial_\rho (D_k S^a)\partial_\sigma A_{i(1)}^0
+\partial_\rho (D_k S)_{(1)}^0\partial_\sigma A_i^a ) \,]\cr &
+\gamma^0\,[ \,e\epsilon^{abc}((D_kS^b)S_{(2)}^c+(D_kS)_{(2)}^0)S^c
-{1\over8}\theta^{\rho\sigma}\theta^{\alpha\beta}
\partial_\rho\partial_\alpha(D_kS^b)\partial_\sigma \partial_\beta S^c
 ) \cr &
\qquad\qquad -{1\over2}{\theta^{\rho\sigma}}(
\partial_\rho (D_k S)_{(1)}^0\partial_\sigma S^a)\,]\,
\}P_+\alpha,\cr
 }}
which vanish because $P_+\alpha=0$ for the broken supersymmetries. The fourth order variations of the 
bosonic fields then vanish because they are each proportional to $\delta^3\psi$. Note, the vanishing of the
third and fourth order variations of the fields implies a vanishing quadrupole moment tensor for all states in
the monopole BPS multiplet and is different from  the variation of the $N=2$ black hole supermultiplet, for which these variations are nonzero \ae. {}  It turns out that the BPS monopole exhibits no electromagnetic quadrupole structure in both the commutative and noncommutative spaces, and that the dipole structure of noncommutative monopole does not give rise to the electric quadrupole moment up to $O(\theta^2)$, the same of which holds apparently for any arbitrary higher order in $\theta$.  This result is disappointing. It is mainly from the fact that noncommutativity influences only to the spacial part of field variations, not to the spin structure when supersymmetry transformation is done. In order to check this, let's see the invariance of the  angular momentum operator.

The fermionic fields $\hat\psi^A$ may be expanded in the monopole background as
\eqn\zeromodes{\eqalign{
\hat\psi^{a\rho}&=-2(\gamma^k)^\rho{}_\sigma\hat\alpha^\sigma D_k\hat S^a 
 + \hbox{nonzero-modes},\cr
\hat\psi^{0\rho}&=-2(\gamma^k)^\rho{}_\sigma\hat\alpha^\sigma 
(D_k\hat S)_{(1)}^0 
+ \hbox{nonzero-modes},\cr
\hat\psi^{a\rho}_{(2)}&=-2(\gamma^k)^\rho{}_\sigma\hat\alpha^\sigma (D_k\hat S)_{(2)}^a) + \hbox{nonzero-modes},\cr
}}
where $\rho,\sigma$ are spinor indices and we have explicitly displayed only the zeromode part of the
expansion. Using the orthogonality of zero-modes and nonzero-modes, we can then express 
the spinorial parameters 
$\hat\alpha^\lambda$ and $\hat\alpha^\dagger_\lambda$ as\foot{see \dka{} for angular momentum operator in detail}
\eqn\transforms{
\hat\alpha^\lambda= +{1\over 2M}
\int d^3x(\gamma^l)^\lambda{}_\rho\hat\psi^{a\rho}D_l\hat S^a,\qquad
\hat\alpha^\dagger_\lambda= -{1\over 2M}
\int d^3x\hat\psi^{a\dagger}_\rho(\gamma^l)^\rho{}_\lambda D_l\hat S^a,}
where $M=4\pi v/e$ is the mass of the monopole\foot{Here we have made use of the result
$\int d^3x\,\eta^{kl}(D_k\hat S^a)D_l\hat S^a = -M$} and this form is that of the commutative case because the mass term arising from noncommutativity 
\eqn\msecond{
\int d^3x\,\eta^{kl}\{(D_k\hat S)_{(2)}^aD_l\hat S^a 
+(D_k\hat S^a)D_l\hat S_{(2)}^a + (D_k\hat S)_{(1)}^0\hat S_{(1)}^0 \} = -M_{(2)}
} 
vanishes\foot{$M_{(1)}$ also vanishes because the scalar solution is not influenced by noncommutativity at $O(\theta)$.}. It is because the contributions from noncommutative fields fall
 off faster than $1\over r^2$ compared to the commutative ones\foot{We use noncommutative BPS solutions in \kha\dba\sgo}, which makes the  second order mass vanish. Consequently, the angular momentum vector has no correction from the noncommutativity as expected,
\eqn\angmom{
J^k=2iM\left (\alpha^\dagger\gamma^{k}\alpha\right).}
%

As an another check, we now turn to the long range limit of the electric field for the monopole superpartner solution up to $O(\theta^2)$.
The result for the long range electric fields ${\hat E}^i=
{\hat F}_{0i}\equiv {1\over v}{\hat S}^A*{\hat F}_{0i}^A$ obtained are
\eqn\dipolefield{\eqalign{ E^i &= -{2i\over e}
\left(\alpha^\dagger\gamma^{k}\alpha\right) 
\left\{{3x^kx^i\over r^5}-{\delta^{ki}\over r^3}\right\},\cr
E_{(1)}^i &= 0, \cr
E_{(2)}^i &= -{2i\over e}\left(\alpha^\dagger\gamma^{k}\alpha\right) 
\left\{\hbox{ no dipole field like terms} \right\}, \cr 
}}
which shows that a dipole field with dipole moment vector $\vec p= -{2i\over e}(\alpha^\dagger\vec\gamma\alpha)$ can be seen only in commutative sector and 
that the electric dipole moment proportional to angular momentum operator, thus also  the gyroelectric ratio $g=2${}{}\liu\osborn\porrati, obtain no corrections from noncommutativity.

In conclusion, we considered $U(2)$ monopole in noncommutative space by constructing superpartner solutions up to $O(\theta^2).$  We found no electric quadrupole moment that is expected\igi {} by the dipole structure of noncommutative $U(2)$ monopole, which is because spin is indpendent of noncommutativity. As a check, 
 up to $O(\theta^2)$ we showed explicitly that the angular momentum operator and the electric dipole moment obtain no correction from noncommutativity. In a more broad perspective, this result might give an example of the nature of how supersymmetry works not changing between the commutative and noncommutative theories.

\bigskip\noindent
{\bf Acknowledgements: }  We thank Dongsu Bak, David Kastor, Kimyeong Lee and Soonkeon Nam for helpful discussions
and correspondence. This work is supported by BK21 Program of Korea Research Foundation.

\listrefs
\end

It is well known that magnetic monopoles satisfying the Dirac quantization condition 
are necessarily the finite configurations with non-zero topological charge in the Yang-Mills-Higgs theory,whose
Lagrangian is 
\eqn\lagrangian{
{\cal L}=-{1\over4}F_{\mu\nu}^a F^{a\mu\nu} +{1\over2}D_{\mu}\Phi^aD_{\mu}\Phi^a-V(\Phi), }
where
\eqn\fields{\eqalign{
F_{\mu\nu}^a&=\partial_{\mu}A_{\nu}^a-\partial_{\nu}A_{\mu}^a +e\epsilon^{abc}A_{\mu}^b A_{\nu}^c,\cr
D_{\mu}\Phi^a&=\partial_{\mu}\Phi^a +e\epsilon^{abc}A_{\mu}^b \Phi^c,\cr}
}
with $a,b,c=1,2,3$ labeling the adjoint representation of $SU(2)$. 
For non-zero vaccuum expectation value of $\Phi$ 
\eqn\potential{
V(\Phi)=\lambda(\Phi^a\Phi^a-v^2)^2/4.
}
The equations of motion are
\eqn\motion{\eqalign{
D_{\mu}F^{\mu\nu}&=e\epsilon^{abc}\Phi^b D^{\nu}\Phi^c, \cr
(D^{\mu}D_{\mu}\Phi)^a&=-\lambda\Phi^a(\Phi^b\Phi^b-v^2), \cr} }
and the Bianchi identity
\eqn\bianchi{
D_{\mu}\ast F^{a\mu\nu}=0.
}

The energy density of field configuration is 
\eqn\energy{
\Theta_{00}={1\over 2}(\vec E^a\cdot\vec E^a +\vec B^a\cdot\vec B^a+\Pi^a\Pi^a
+\vec D\Phi^a\cdot\vec D\Phi^a)+V(\Phi), }
where $\Pi^a$ is the conjugate momentum of field $\Phi$,$\Pi^a=D_0\Phi^a$ and 
non-abelian electric and magnetic fields are 
\eqn\nonabelian{E^{ai}=-F^{a0i},\qquad B^{ai}=-{1\over2}\epsilon^{ijk}F_{jk}^a. }
The vacuum is given by a configuration with vanishing gauge field, with a constant Higgs filed $\Phi^a$
\eqn\vacuum{
F^{a\mu\nu}=D^\mu \Phi^a=V(\Phi)=0,\qquad \Phi^a\Phi^a=v^2.}
A constant Higgs field breaks $SU(2)$ gauge symmetry to $U(1)$ subgroup.

Magnetic charge is defined as
\eqn\magnetic{
q={1\over4\pi}\int_{S_{\infty}^2}\vec B\cdot d\vec S ={1\over4\pi v}
\int_{S_{\infty}^2}\Phi^a\vec B^a\cdot d\vec S ={1\over4\pi v}\int\vec B^a\cdot (\vec D\Phi)^a d^3r, 
}
using the Bianchi identity and integration by parts. 
The energy of the static configuration with vanishing electric field is given by
\eqn\static{\eqalign{
M_M&=\int d^3r\left({1\over2}(\vec B^a\cdot\vec B^a +\vec D\Phi^a\cdot\vec D\Phi^a)+V(\Phi)\right)\cr
&\geq{1\over2}\int d^3r(\vec B^a-\vec D\Phi^a)\cdot (\vec B^a-\vec D\Phi^a)+4\pi vq,\cr}
}
using \magnetic. From this we have Bogomol'nyi bound 
\eqn\bogo{M_M\geq 4\pi vq. }
and two conditions
\eqn\conditions{V(\Phi)=0,\qquad \vec B^a=\vec D\Phi^a, }
which saturate the bound. The second condition is called as Bogomol'nyi equation. 
In supersymmetric theories which have potentials with exact flat directions protected by supersymmetry, symmetry
breaking makes sense with the first condition if we impose a boundary condition 
\eqn\boundary{ \Phi^a\Phi^a\to v^2\;\;\;{\rm as}\;\, r\to\infty, } 
for arbitrary v.

A simple solution to the equation of motion with Bogomol'nyi condition is 
found by Prasad-Sommerfield in terms of two radial functions $H,\,K$
\eqn\prasad{\eqalign{
\Phi^a&={\hat r^a\over er}H(ver),\cr
A_i^a&=-\epsilon_{ij}^a{{\hat r}^j\over er}(1-K(ver)).\cr} }
with the boundary conditions
\eqn\conditions{\eqalign{
K(ver)&\rightarrow 1,\qquad H(ver)\rightarrow 0,\quad {\rm as}\;r\rightarrow 0,\cr 
K(ver)&\rightarrow  0,\qquad H(ver)/(ver)\rightarrow 1,\quad {\rm as}\quad 
r\rightarrow \infty.\cr }}
where
\eqn\functions{
H(y)=y\coth y-1, \qquad
K(y)={y\over\sinh y}.}
For large $r$ we thus have
\eqn\asymptotic{
\Phi^a\rightarrow v\hat r^a-{\hat r^a\over er} A_i^a\rightarrow -\epsilon_{ij}^a{{\hat r}^j\over er}.} 

\newsec{Monopoles in N=2 Supersymmetric Gauge Theory} 
The Lagrangian of $N=2$ Super Yang-Mills theory is given by 
\eqn\super{\eqalign{
{\cal L}_{N=2}&={\rm Tr}(-{1\over4}F_{\mu\nu}F^{\mu\nu}
-{1\over4}(D_{\mu}P)^2-{1\over2}(D_{\mu}S)^2-{e^2\over2}[S,P]^2 \cr &
+i\bar\psi\gamma^{\mu}D_{\mu}\psi-e\bar\psi[S,\psi] -e\bar\psi\gamma_5[P,\psi]) \cr} }
where all fields
are elements of $SU(2)$ Lie algebra, {\it i.e.} $S=S^a T^a$ {\it etc.} and $S$ and $P$ are scalar Higgs fields.
This Lagrangian is invariant under the supersymmetry transformations 

\eqn\susyvariations{\eqalign{
\delta A_{\mu}&=i\bar\alpha\gamma_{\mu}\psi -i\bar\psi\gamma_{\mu}\alpha, \cr
\delta P&=\bar\alpha\gamma_5\psi
-\bar\psi\gamma_5\alpha, \cr
\delta S&=i\bar\alpha\psi
-i\bar\psi\alpha, \cr
\delta\psi&=(\sigma^{\mu\nu}F_{\mu\nu}-i\gamma^\mu D_\mu S +\gamma^\mu D_\mu P\gamma_5-i[P,S]\gamma_5)\alpha, \cr }}
with $\alpha$ the supersymmetry parameter of the grassmann valued Dirac spinor 
which is equivalently two Majorana spinors. 

We are now going to find the electric dipole moment of magnetic monopole doing higher order 
supersymmetry variation of fields. All the fields are considered as
\eqn\taylorseries{
X_\mu=X_{\mu}^{(0)}+X_{\mu}^{(1)}+{1\over2}X_{\mu}^{(2)}+\dots }
where $X_{\mu}^{(n)}=\delta\,.\,.\,.\,\delta X_{\mu}$ with $n\,\delta$s. 
We start with a classical solution with the fermion field set to zero,choose gauge condition $A_0=0$ and set
$P^a=0$ by chiral rotation. Our boundary condition is as before $S^a S^a \rightarrow v^2$ as $r\rightarrow\infty$
and the Bogomol'nyi equation becomes 
\eqn\susybogo{
B_i=D_i S =-{1\over2}\epsilon_{ijk}F^{jk}. }
Then we have
\eqn\firstorder{\eqalign{
\psi^{(1)}&=\delta\psi=(\sigma^{\mu\nu}F_{\mu\nu}-\rlap/{D}S)\alpha\cr 
&=-\gamma^i B_i(1-\Gamma_5)\alpha\cr
&=-2(\gamma^i D_iS)\alpha_-. \cr} }
Our conventions and definitions used above are 
\eqn\conventions{
\epsilon^{0123}=+1,\qquad \eta^{\mu\nu}=(1,-1,-1,-1), }
and
\eqn\moreconventions{\eqalign{
\{\gamma^{\mu},\gamma^{\nu}\}&=2g^{\mu\nu},\qquad \sigma^{\mu\nu}={1\over4}[\gamma^{\mu},\gamma^{\nu}],\cr
\gamma_5&=i\gamma^0\gamma^1\gamma^3\gamma^3,\qquad \Gamma_5=i\gamma^0\gamma_5 \cr
\Gamma_{\pm}&={1\over2}(1\pm\Gamma_5),\qquad \alpha_{\pm}=\Gamma_{\pm}\alpha.\cr} }
where the supersymmetries $\alpha_+$ are unbroken in the monopole background while the $\alpha_-$ 
supersymmetries are broken,that is $\alpha_+=0$.
Equation \firstorder\ for the broken supersymmetries give zero energy Grassmann 
variations of the monopole solution.
Thus we have
\eqn\allfirst{\eqalign{
\bar\psi^{(1)}&=-2\bar\alpha_-(\gamma^i D_i S),\cr
 A_{\mu}^{(1)}&=0,\cr
P^{(1)}&=0,\cr
S^{(1)}&=0\cr}}
The second order variations are by using the relation 
\eqn\trick{
\gamma_5\alpha_-=i\gamma^0\alpha_-,
}
\eqn\secondorder{\eqalign{
A_{0}^{(2)}&=P^{(2)}
=-4i(D_i S)(\bar\alpha_-\gamma_0\gamma^i\alpha_-), \cr 
A_{i}^{(2)}&=0=S^{(2)}=\psi^{(2)}=\bar\psi^{(2)}. \cr}}
Unexpectedly all the higher order variations are identically zero. 

Now we can write the component of the electric field arising from moving magnetic charge of monopole as 
\eqn\electric{
E^{ia}=\partial_i{A_0^{(2)a}\over2}
-e\epsilon^{abc}{A_0^{(2)b}\over2}A_i^{(0)c}, }
and by using equation \asymptotic\ and the finite energy condition 
\eqn\abelianpart{
E_i=E_i^a{S^a\over v},
}
with
\eqn\anothertrick{
-i\gamma_i\gamma_0\gamma_5={1\over4}\epsilon_{ijk}\gamma^{jk},\qquad \gamma_{kl}\equiv [\gamma_k,\gamma_l],
}
we have
\eqn\finally{\eqalign{
E_i&={i\over2}\epsilon^{jkl}(\alpha_-^{\dagger}\gamma_{kl}\alpha_-) \left({1\over
ver}-1\right)\left({\eta_{ij}\over e r^3} -{3x_i x_j\over e r^5}\right)\cr 
&=\left({3x_i x_j\over
r^5}-{\eta_{ij}\over e r^3}\right) {\cal P}^j+\dots\cr} }
where
\eqn\operator{
{\cal P}^j={i\over2e}\epsilon^{jkl}\alpha_-^{\dagger}\gamma_{kl}\alpha_-, }
denotes the electric dipole moment of moving magnetic monopole.
Since the angular momentum of spin ${1\over2}$ fermion is defined as 

\eqn\angmom{
J^{\mu\alpha\beta}=x^\alpha\Theta^{\mu\beta} -x^\beta\Theta^{\mu\alpha}
+{1\over2}\bar\psi\{\gamma^{\mu},{i\over 4}\gamma^{\alpha\beta}\}\psi, }
where $\Theta^{\mu\nu}$ is the energy-momentum tensor density. 
The purely spin part which is arising from the last term is written 

\eqn\spinangmom{
J^{0jk}={1\over2}\bar\psi^{(1)}
\{\gamma^{0},{i\over4}\gamma^{\alpha\beta}\}\psi^{(1)}, }

and thus
\eqn\result{\eqalign{
\epsilon_{ijk}J^{0jk}&=-{i\over2}\epsilon_{ijk}\alpha_-^{\dagger} \gamma^{jk}\alpha_-D^l S D_l S\cr
&=-e{\cal P}^i D^l S D_l S,\cr
}}
is straight forward.
Therefore spin component ${\cal S}^i$ is 

\eqn\spinvector{\eqalign{
{\cal S}^i&=\int{\rm d}^3x\epsilon^{ijk}J_{0jk}\cr
&=-e{\cal P}^i\int{\rm d}^3x\,\partial_l(S^a D^l S^a)\cr
&=-e{\cal P}^i\int{\rm d}\theta{\rm d}\varphi\,{\rm sin}\theta\, r x_l(S^a D^l S^a)\cr 
&=-4\pi e{\cal P}^i\left[-{r\over e}({v\over r}-{1\over er^2}) \right]_{r\rightarrow\infty}\cr 
&=4\pi v{\cal P}^i \cr}}
where we use equations \motion\ and \asymptotic\ and

\eqn\yetanothertrick{
D_l(S^a D^l S^a)=\partial_l(S^a D^l S^a), }
hence the electric dipole moment of monopole ${\cal P}^i$ is 

\eqn\dipolemoment{
{\cal P}^i={{\cal S}^i\over4\pi v}\equiv{g_e q\over2M_M}{\cal S}^i,}
where $g_e$ is the gyroelectric ratio and finally using equation \bogo\ we have

\eqn\gyroelectric{
g_e=2.
}

\newsec{The Electric Dipole Moment}

\listrefs
\end

\newcommand{\eqn{} {\eqn{gin{equation}}
\newcommand{}} {\end{equation}}
\newcommand{\eqn{a} {\eqn{gin{eqnarray}}
\newcommand{}a} {\end{eqnarray}}

\newcommand{\ba} {\eqn{gin{array}}
\newcommand{\ea} {\end{array}}
\newcommand{\cr} {\noindent}
\newcommand{\cr} {\nonumber \\}
\newcommand{\no} {\nonumber}
\newcommand{\ul} {\underline}
\newcommand{\wt} {\widetilde}
\newcommand{\wh} {\widehat}
\newcommand{\bfl} {\eqn{gin{flushleft}}
\newcommand{\efl} {\end{flushleft}}